\documentclass[superscriptaddress, preprint, amsmath, amssymb, aps, prb, floatfix, nofootinbib]{revtex4-2}
\usepackage[utf8x]{inputenc}
\usepackage{amsmath}\usepackage{xcolor}%
\usepackage{graphicx}% Include figure files
\usepackage{dcolumn}% Align table columns on decimal point
\usepackage{bm}% bold math
\usepackage{subfigure}
\usepackage{textcomp}

\newcommand{\ie}{\emph{i.e.}, }

\newcommand{\nn}{\nonumber}
\newcommand{\nnl}{\\\nonumber}

\newcommand{\ket}[1]{\left| #1 \right\rangle}
\newcommand{\bra}[1]{\left\langle #1 \right|}
\newcommand{\iprod}[2]{\left\langle #1 \big\vert #2 \right\rangle}

\newcommand{\abs}[1]{\left| #1 \right|}
\newcommand{\dpar}[2]{\frac{\partial #1}{\partial #2}}

\newcommand{\expn}[1]{{\rm e}^{#1}}
\newcommand{\bv}[1]{\mathbf{#1}}
\newcommand{\dg}{{^{\dagger}}}

\newcommand{\bvk}{\bv{k}}
\newcommand{\bvq}{\bv{q}}
\newcommand{\bvr}{\bv{r}}

\newcommand{\dl}{$\delta$-layer}
\newcommand{\cmi}{cm$^{-1}$}
\newcommand{\Imt}{{\rm Im}}

\begin{document}
	\title{Suppression of mid-infrared plasma resonance due to quantum confinement in delta-doped silicon}

\author{Steve M. Young}

\author{Aaron M. Katzenmeyer}

\author{Evan M. Anderson}
\author{Ting S. Luk}
\author{Jeffrey A. Ivie}
\author{Scott W. Schmucker}
\author{Xujiao Gao}
\author{Shashank Misra}

\affiliation{ Sandia National Laboratories, Albuquerque, New Mexico 87185, USA}

\date{\today}

\begin{abstract}

The classical Drude model provides an accurate description of the plasma resonance of three-dimensional materials, but only partially explains two-dimensional systems where quantum mechanical effects dominate such as P:\dl s  -- atomically thin sheets of phosphorus dopants in silicon that induce novel electronic properties beyond traditional doping. Previously it was shown that P:\dl s  produce a distinct Drude tail feature in ellipsometry measurements.    However, the ellipsometric spectra could not be properly fit by modeling the \dl~as discrete layer of classical Drude metal. In particular, even for large broadening corresponding to extremely short relaxation times, a plasma resonance feature was anticipated but not evident in the experimental data. In this work, we develop a physically accurate description of this system, which reveals a general approach to designing thin films with intentionally suppressed plasma resonances.  Our model takes into account the strong charge density confinement and resulting quantum mechanical description of a P:\dl.  We show that the absence of a plasma resonance feature results from a combination of two factors: i), the sharply varying charge density profile due to strong confinement in the direction of growth; and ii), the effective mass and relaxation time anisotropy due to valley degeneracy.  The plasma resonance reappears when the atoms composing the \dl~are allowed to diffuse out from the plane of the layer, destroying its well-confined two-dimensional character that is critical to its novel electronic properties.
\end{abstract}
\maketitle
\section{Introduction}
The plasma resonance is a fundamental property of the AC response of a material - the frequency where its dielectric function changes sign. While this was first explained in the context of the classical scattering of electrons by Drude, the plasma resonance survives in more complicated circumstances, ranging from the two-dimensional limit \cite{Shah_2022} to fully quantum mechanical treatments \cite{Mendoza_2021}. Indeed, the suppression of the plasma resonance in  Au/CdS nanocomposites came as a surprise \cite{Khon_2011}, and is a consequence of strong mixing between the states of the metal nanoparticle and the semiconducting nanorod. Here, we determine that the suppression of the plasma resonance in a single-component material system – a two-dimensional sheet of doped silicon – arises from strong anisotropy of the electronic structure in-plane versus out-of-plane. 

Atomic Precision Advanced Manufacturing (APAM) leverages site-selective surface chemistry to incorporate dopant atoms into two-dimensional sheets in silicon (\dl s) with lateral precision down to the single-atom scale when patterned with scanning tunneling microscopes on one hand, and density that exceeds the solid solubility limit on the other, enabling the fabrication of novel devices. Single-atom precision has been used in the fabrication of high-quality quantum bit devices \cite{Buch_2013,He_2019,Watson_2017}, and sub-nm precision for analog simulation of quantum materials concepts\cite{Wang_2022,Kiczynski_2022}. The exceptionally high dopant density produces strong confinement and a novel electronic structure in silicon \cite{Mazzola2020}, and has led to exploration of new types of classical electronic devices \cite{Weber_2012,Ward2020,Tzu-Ming_2021,Mendez_2022}. Characterization of their optical response using infrared-variable angle spectroscopic ellipsometry (IR-VASE) has shown the clear influence of an APAM phosphorus \dl~ on the mid- and long-wavelength infrared optical response of silicon, as measured by the change in the reflected light’s polarization \cite{Katzenmeyer2020}. However, modeling the \dl~ with the complex permittivity of silicon modified by a Drude oscillator creates a plasma resonance, observed in ion implanted silicon \cite{Ginn2011}, but never in APAM material over an extended range of wavelengths. While the broad tail could be well-fit by the Drude model, the experimental absence of the predicted plasma resonance calls into question the accuracy of the model and extracted sample parameters. 

Thus, developing a new model of the optical response of APAM materials is necessary to fully understand the unique optical properties of ultra-doped silicon, helping to define their potential for silicon optical devices. Here we show that the calculation of the electromagnetic response using a quantum mechanical description of a layer of free charge within a dielectric stack correctly reproduces the ellipsometric data from the \dl. With the revised model, we explain how the lack of a plasma resonance emerges as a consequence of confinement on the out-of-plane wavefunctions and the anisotropic carrier effective masses and relaxation times. We show the dependence of spectral features on dopant layer properties, such as electronic width of the \dl~and relaxation, including the re-emergence of a plasma resonance in sufficiently diffused layers. Practically, the combination of ellipsometry and this new theoretical approach provides a straightforward, non-destructive way to characterize the thickness of a \dl. More broadly, this work identifies a general approach to suppressing, e.g. undesirable, plasma resonances in thin films through careful choice of electronic structure and dimensionality.

\section{Modeling}

Spectroscopic ellipsometric data depend on the polarization dependent reflectivity response of a material, and require a suitable model for the material's optical response in order to properly interpret the results and understand the properties of its constituents.  Typically, heterogeneous media can be represented as stacks of discrete homogeneous layers characterized by their bulk dielectric behavior.  This approach works well, even for very layers down to a few nanometers or less~\cite{Woollam1990,Tiwald1998,Yoo2022}.  However, in the present case we are considering potentially atomically thin materials that behave qualitatively differently from thin but nonetheless bulk 3D materials.  To address this we will refine the conventional approach in two ways. First, we treat the \dl~as a defect layer within a dielectric slab, rather than a distinct material layer~\cite{Jablan_2009,Li_2018}.  As such, it will be described by a spatially-dependent conductivity $\sigma(\bvq,z)$, where $z$ is the growth direction and $\bvq$ the in-plane wavevector, rather than a finite region associated with a spatially uniform permittivity function $\epsilon(\bvq)$~\cite{Cleary2010}.  Second, while the ellipsometry spectra will be computed classically from Maxwell's equations, the function $\sigma(\bvq,z)$ will be determined from a quantum mechanical description of the \dl, based on its observed electronic structure (Fig.~\ref{fig:bands}).

\subsection{P:\dl~Electronic Structure}
\begin{figure}
	\centering
	\subfigure[]{\includegraphics[width=.25\columnwidth]{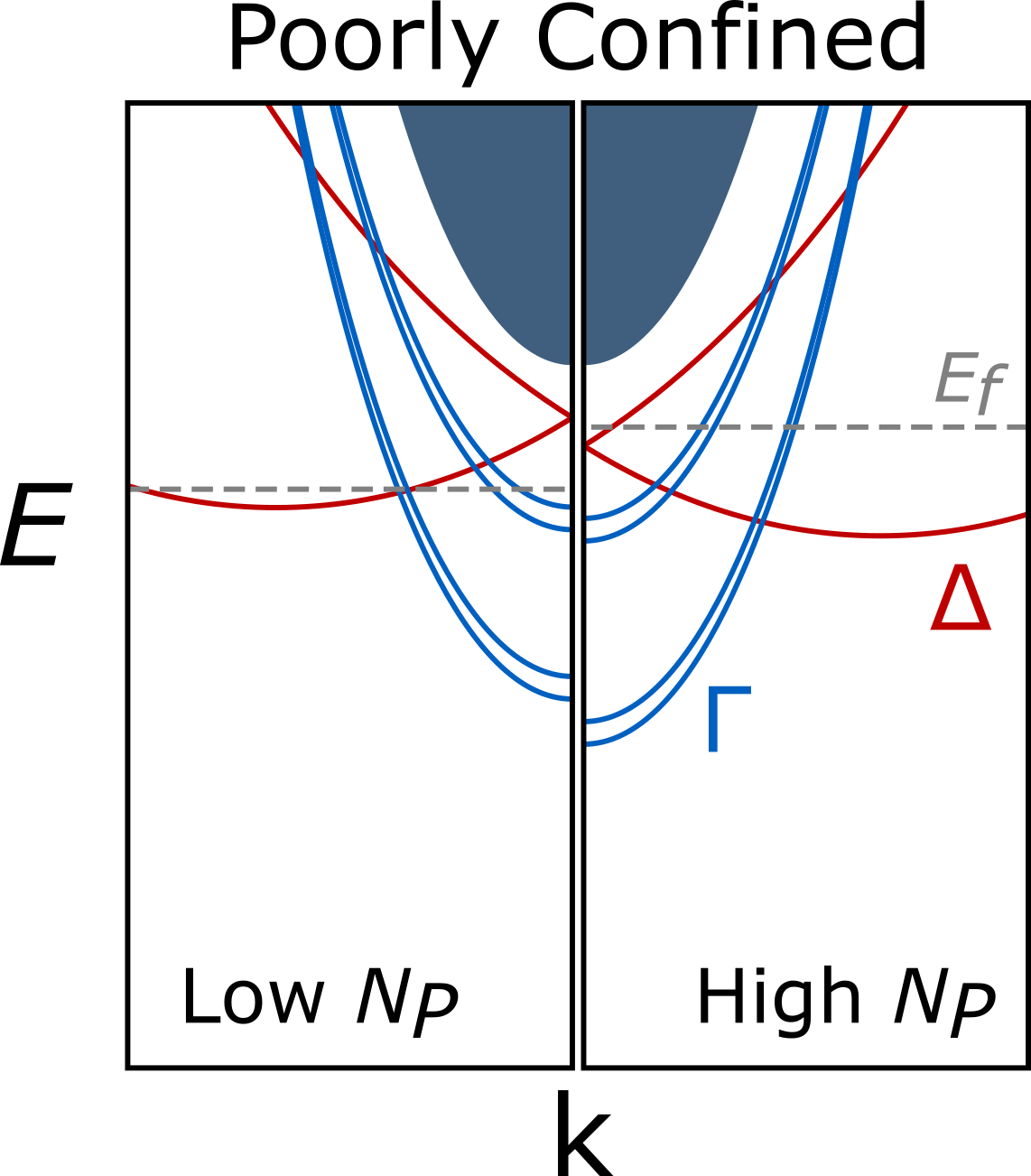}\label{fig:bands1}}
	\subfigure[]{\includegraphics[width=.25\columnwidth]{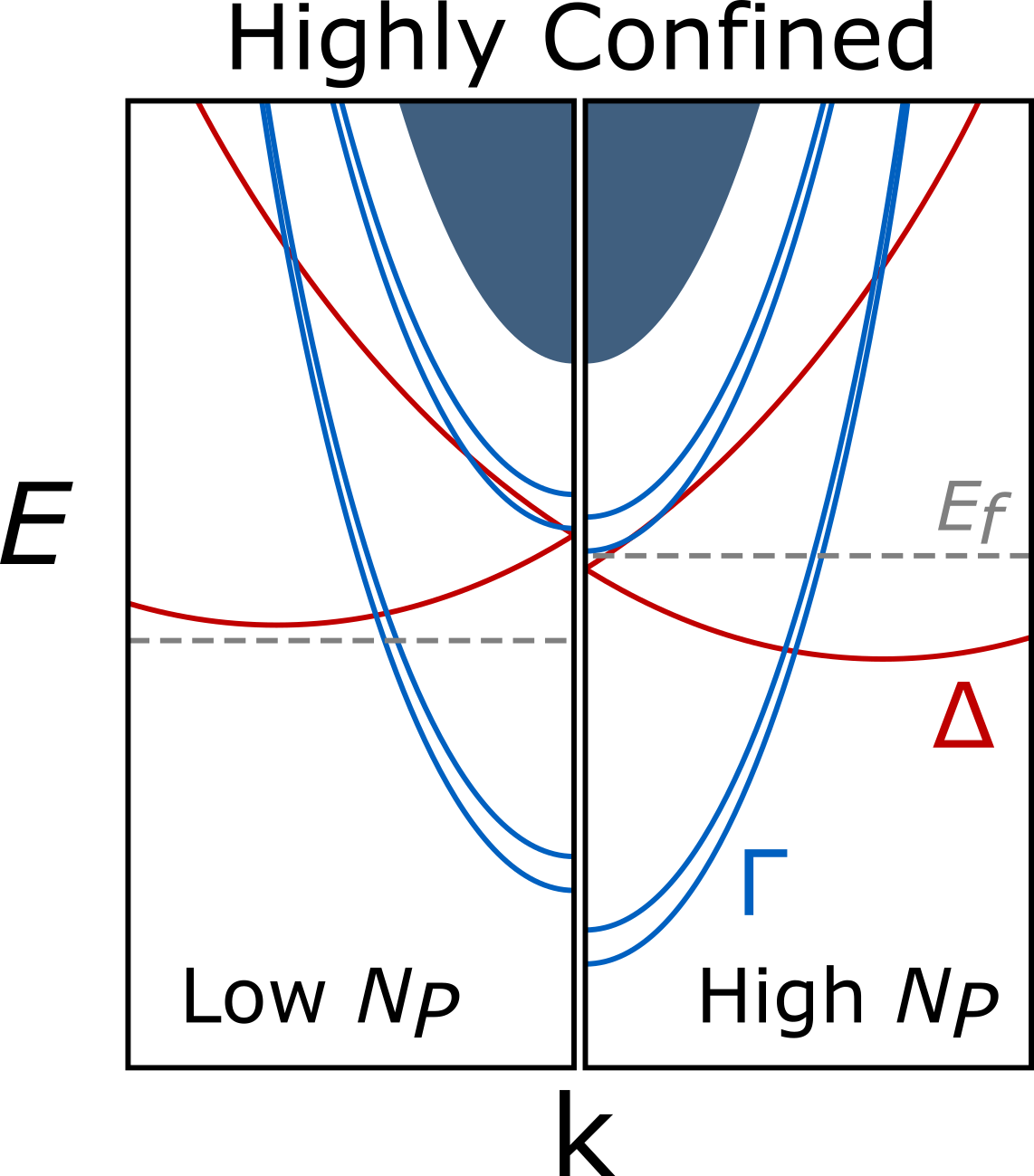}\label{fig:bands2}}
	\caption{ Cartoon schematics of typical band structures of phosphorus \dl s in silicon~\cite{Lee2011,Carter2011,Mazzola2020,Holt2020} highlighting the qualitative features presently of interest. The \dl, characterized by a areal density of phosphorous atoms $N_p$, generates a confining potential which creates localized bands crossed by a Fermi energy below the silicon conduction band. As shown experimentally\cite{Holt2020}, the density and confinement of phosphorus atoms together control both the band splitting and the type of carriers composing the electronic system.  In poorly confined \dl s~\subref{fig:bands1}, the energy splitting of the bands is relatively weak, so that $\Delta$ bands (red), originating from the transverse valleys in silicon, are populated by carriers, along with multiple pairs of $\Gamma$ bands (blue) originating from the two longitudinal valleys, even for lower areal densities of phosphorous. In highly confined layers~\subref{fig:bands2}, the gap between subbands is larger, so that carriers populate a single pair of $\Gamma$ bands and occupy $\Delta$ bands at only higher areal densities. \label{fig:bands}}
\end{figure}

The conduction band of silicon comprises six degenerate valleys; substitution of silicon with phosphorus atoms populates the conduction band with electrons. A sufficiently dense, uniform phosphorus \dl~breaks the translational symmetry in the direction perpendicular to the layer, creating confined states that manifest as lower-in-energy subbands.  Cartoon illustrations of the resulting two-dimensional (2D) band structure are shown in Fig.~\ref{fig:bands}.   Bands arising from the longitudinal valleys appear at the $\Gamma$ points and those originating from the transverse valleys appear away from $\Gamma$ and are commonly labeled as $\Delta$ bands.    The $\Gamma$ subbands come in nearly degnerate pairs (one from each longitudinal valley) -- usually split by a few tens of meV -- with each pair separated by $>200$meV for high dopant densities/confinements.  The $\Delta$ subbands appear higher in energy, though their location varies considerably depending on the density and spatial distribution of dopants~\cite{Lee2011,Carter2011}.  In more diffuse layers, the $\Delta$ and $\Gamma$ bands are closer in energy and carriers will generally populate both (Fig.\ref{fig:bands1}).  On the other hand, for atomically thin \dl s,  most carriers will reside in the lowest pair of $\Gamma$ bands (Fig.\ref{fig:bands2})~\cite{Mazzola2020,Holt2020}, with high densities required to begin filling the $\Delta$ bands.

\subsection{\dl~AC Conductivity}
We model the \dl~as resulting from a vertically oriented confining potential against a homogeneous background within the effective mass approximation~\cite{Ando1982}. It is important to note that we are interested in the full 3D analysis of system physics. While the translational invariance prevails in the plane of the layer, it is broken in the perpendicular ($\hat{z}$) direction; we will represent the in-plane degrees of freedom in momentum space, while retaining the $\hat{z}$ coordinate in real space.  The eigenstates and energies of the non-interacting Hamiltonian $H_v^{0}=H_v^{0,\parallel}+H_v^{0,\perp}$ are then defined as
\begin{flalign}
	\psi_{vn}(\bvk,z)&=\expn{i\bvk\cdot \bvr}\psi_{vn}(z)=\expn{i\bvk\cdot \bvr}\iprod{z}{v,n}\nn\\
	E_{vn}(\bvk)&=\varepsilon_{vn}+\frac{1}{2}\left(\frac{k_x^2}{ m_{vx}}+\frac{k_y^2}{ m_{vy}}\right)\label{eq:wfs}
\end{flalign}

where $n$ indexes the eigenstates of $H^{0,\perp}$, $v$ indexes the valleys $\Gamma$ and $\Delta$, and $m_{vx}$, $m_{vy}$ are the in-plane effective masses for valley $v$, which may be silicon's transverse effective mass $m_\perp$ or longitudinal effective mass $m_\parallel$ depending on the valley.	Coupling between valleys is relatively small, including the coupling between the two orginally degenerate $\Gamma$ valleys, and will be taken to be zero for simplicity.  For convenience we define in each valley $\ket{z}=\vec{\psi}_v(\bvk,z)=\sum_n\psi_{vn}(\bvk,z)\dg\ket{v,n}$.

Within the Kubo-Greenwood formalism for a non-interacting system,	the two-point, frequency-dependent conductivity tensor $\sigma_{ij}$ may be written as 
\begin{flalign}
	\sigma_{ij}(\omega,\bvr,\bvr')=ie^2\sum_{mn}\frac{n_F(E_n)-n_F(E_m)}{E_n-E_m}\frac{\bra{m}\mathcal{J}_i(z)\ket{n}\bra{n}\mathcal{J}_j(z')\ket{m}}{\omega+E_n-E_m+\frac{i}{\tau}}\label{eq:kg}
\end{flalign}
where $n_F(E)$ is the Fermi function, $\hat{\mathcal{J}}(z)$ is the current density operator, $n$ and $m$ index states of the system, and  $\tau$ is infinitely large.  

The sum over states can be broken into sums over valleys ($v$), in-plane wavevectors ($\bvk$), and bands ($n$, $m$), and the expression can rewritten in terms of a wavevector ($\bvq$) for the in-plane degrees of freedom rather than coordinates as 
\begin{flalign*}
	\sigma_{ij}(\omega,\bvq,z,z')=ie^2\sum_{vmn}\int \frac{d\bvk}{(2\pi)^2}&\frac{n_F(E_{vn}(\bvk))-n_F(E_{vm}(\bvk+\bvq))}{E_{vn}(\bvk)-E_{vm}(\bvk+\bvq)}\\
	&\times\frac{\bra{m,\bvk+\bvq}\mathcal{J}_i(z)\ket{n,\bvk}\bra{n,\bvk}\mathcal{J}_j(z')\ket{m,\bvk+\bvq}}{\omega+E_{vn}(\bvk)-E_{vm}(\bvk+\bvq)+\frac{i}{\tau_v}}
\end{flalign*}	
where $z$, $z'$ are out-of-plane coordinates. In what follows we will incorporate interactions and many-body effects by treating $\tau$ as finite and roughly corresponding to the relaxation time in the \dl.  While this violates particle conservation\cite{Baym1962,Mermin1970,abrikosov2012} in general, in the long-wavelength limit that is relevant here it is consistent with the corrected expression of \cite{Mermin1970} and yields an appropriate expression for our purposes\cite{Vos_2017}. We note that there are distinct relaxation times for the two valleys, $\tau_\Gamma$ and $\tau_\Delta$.  As shown in \cite{Hwang2013}, the relaxation time derived from the expression for the self-energy includes a dependence on the density of states and Fermi wavevector $k^F$ of the bands involved in scattering for a given impurity potential as $\tau\propto \frac{(k^F)^2}{m}$. This is important since we are considering valleys with different electronic structure; while this factor is normally absorbed into the overall expression for the relaxation time and need not be considered explicitly, it must appear in our expression.  Specifically, we will write the relaxation times in terms of a baseline $\tau$, so that $\tau_\Gamma=\tau$ and $\tau_\Delta=\sqrt{\frac{m_\perp}{m_\parallel}}\left(\frac{k^F_\Delta}{k^F_\Gamma}\right)^2\tau$. Taking the long-wavelength limit $\bvq\rightarrow0$
\begin{flalign}
	\sigma_{xx}(\omega,z,z')&=\sum_{vn}\int \frac{d\bvk}{(2\pi)^2}\dpar{n_F}{E_{vn}}\frac{k_x^2}{m_{vx}^2}\frac{\abs{\psi_{vn}(z)}^2\abs{\psi_{vn}(z')}^2}{\omega+\frac{i}{\tau_v}}\nnl
	\sigma_{zz}(\omega,z,z')&=\sum_{v,n\ne m}\int \frac{d\bvk}{(2\pi)^2}\frac{n_F(E_{vn}(\bvk))-n_F(E_{vm}(\bvk))}{E_{vn}(\bvk)-E_{vm}(\bvk)}\nnl
	&\times\frac{1}{m_{vz}^2}\frac{\left(\psi_{vn}\dpar{\psi_{vm}^*}{z}-\psi_{vm}^*\dpar{\psi_{vn}}{z}\right)\left(\psi_{vm}\dpar{\psi_{vn}^*}{z'}-\psi_{vn}^*\dpar{\psi_{vm}}{z'}\right)}{\omega+E_{vn}(\bvk)-E_{vm}(\bvk)+\frac{i}{\tau_v}}\label{eq:sigma_cond}
\end{flalign}	
where we have used the fact that 
\begin{flalign}
	\bra{m,\bvk}\mathcal{J}_i(z)\ket{n,\bvk}=\left\{\delta_{nm}\frac{k_x\abs{\psi_{vn}(z)}^2}{m_{vx}},\delta_{nm}\frac{k_y\abs{\psi_{vn}(z)}^2}{m_{vy}},\frac{\left(\psi_{vn}\dpar{\psi_{vm}^*}{z}-\psi_{vm}^*\dpar{\psi_{vn}}{z}\right)}{m_{vz}}\right\}.
\end{flalign}
Since we expect the wavefunctions to be very narrow compared to the variation in the electric fields under consideration we will integrate over the second out-of-plane coordinate to yield a local conductivity function.  Additionally, in order to obtain a simple form for the conductivity, we will assume that
the out-of-plane wavefunctions $\psi_{vn}(z)$ can be approximated as harmonic oscillator eigenstates with energies $\varepsilon_{vn}=\omega_{v}(n-1/2)$.  Under these assumptions, the above becomes
\begin{flalign*}
	\sigma_{xx}(\omega,z)&=e^2n_{2D}\sum_{vn}\frac{\chi_{v n}}{m_{vx}}\frac{\abs{\psi_{v n}(z)}^2}{\omega+\frac{i}{\tau_v}}\\
	\sigma_{zz}(\omega,z)&=e^2n_{2D}\sum_{vn}\frac{\chi_{v n}}{m_{vz}}\abs{\psi_{v n}(z)}^2\frac{\omega+\frac{i}{\tau_v}}{(\omega+\frac{i}{\tau_v})^2-\omega_v^2}
\end{flalign*}	
where $\chi_{vn}$ are the fractions of the total density in the indexed state. 

We consider two limits of interest.  First, the conventional case when confinement is poor so that the system can be described essentially as isotropic and the electronic structure is essentially bulk-like, so that $\omega_\Delta$ and $\omega_\Gamma$ are small approaching zero, and $\tau_\Delta=\tau_\Gamma$.   We then have the conventional Drude model
\begin{flalign}
	\sigma(\omega,z)=&\frac{ie^2}{m_{\rm Si}}\frac{f(z)}{\omega+ i\frac{1}{\tau}}\left[\begin{array}{ccc}
		1 & 0 & 0\\
		0 & 1 & 0\\
		0 & 0 & 1\\
	\end{array}\right]\label{eq:sig1}
\end{flalign}
with the density modulated by a smooth envelope function $f(z)$.  Unless otherwise noted, take $f(z)$ to be Gaussian with its width $w_{\rm DL}$ corresponding to the standard deviation and taken as a free parameter.  

Second, to represent \dl s with high carrier densities~\cite{Holt2020} in which only the lowest pairs of bands in each valley sector are occupied, we take $\chi_\Delta=\chi_\Gamma$ and obtain
\begin{flalign}
	\sigma(\omega,z)=&ie^2N(z)\left\{\frac{1}{2}\frac{\abs{\psi_{\Gamma}(z)}^2}{\omega+ i\frac{1}{\tau_\Gamma}}\left[\begin{array}{ccc}
		\frac{1}{m_\perp} & 0& 0\\
		0 &\frac{1}{m_\perp}& 0\\
		0 & 0 &\frac{1}{m_\parallel}\frac{\left(\omega+ i\frac{1}{\tau}\right)^2}{\left(\omega+ i\frac{1}{\tau}\right)^2-\omega_\Gamma^2}\end{array}\right]\right.\nnl
	&
	+\left.\frac{1}{2}\frac{\abs{\psi_{\Delta}(z)}^2}{\omega+ i\frac{1}{\tau_\Delta}}\left[\begin{array}{ccc}
		\frac{1}{2} \left(\frac{1}{m_\perp}+\frac{1}{m_\parallel}\right)& 0& 0\\
		0 &\frac{1}{2}\left(\frac{1}{m_\perp}+\frac{1}{m_\parallel}\right)& 0\\
		0 & 0 &\frac{1}{m_\perp}\frac{\left(\omega+ i\frac{1}{\tau_\Delta}\right)^2}{\left(\omega+ i\frac{1}{\tau_\Delta}\right)^2-\omega_\Delta^2}\\
	\end{array}\right]\right\}\label{eq:sig2}
\end{flalign}
This model is anisotropic due to both the difference in longitudinal and transverse effective masses and the confinement introducing the transition frequency $\omega_\Gamma$ into the out-of-plane response. The standard deviations of the Gaussian density contributions $\abs{\psi_{\Gamma 1}(z)}^2$ and $\abs{\psi_{\Delta 1}(z)}^2$ are naturally determined by $\omega_\Gamma$ and $\omega_\Delta$; we will consider the associated width parameter to be $w_{\rm DL}=\frac{1}{\sqrt{2m_\parallel\omega_\Gamma}}$. Approximating $k^F_\Delta$ as that of an isotropic band with $m=\sqrt{m_\perp m_\parallel}$ and using the condition $\chi_\Delta=\chi_\Gamma$, we find that $\left(\frac{k^F_\Gamma}{k^F_\Delta}\right)^2=2$ and $\tau_\Delta\approx4.5\tau$. We will consider these two models to assess the impact of electronic structure on ellipsometry spectra; specifically Model I will represent poor-quality case where the energy spacing of bands is small so that many bands in each valley are occupied, while Model II will represent the high quality case where only the lowest lying bands are occupied and the system is properly a \dl.

\subsection{Ellipsometry}
For ellipsometry, we are interested in the spectra
\begin{flalign*}
	\Psi(\omega)&=\tan^{-1}\abs{\frac{r_p}{r_s}}\\
	\Delta(\omega)&=\Imt \ln\left(\frac{r_p}{r_s}\right)
\end{flalign*}
so that 
\begin{flalign*}
	\frac{r_p}{r_s}&=\tan\left(\Psi(\omega)\right) \expn{i\Delta(\omega)}
\end{flalign*}
where $r_s$ and $r_p$ are the reflection coefficients of $s$-polarized and  $p$-polarized incident light, respectively. Since these quantities correspond to the magnitude ($\Psi$) and phase ($\Delta$) of the ratio of the reflection coefficients, they contain information about the dielectric behavior of a material but are insensitive to the strength of the probing field.  To simulate these spectra requires that we solve Maxwell's equations for a sample described by $\sigma(\omega,z)$ to obtain the reflection coefficients.  
For the solution to Maxwell's equations we generalize the conventional transfer matrix approach to continuously varying media.  Assuming in-plane translational invariance, we have 
\begin{flalign*}
	\left(i\bvq+\hat{\bv{z}}\dpar{\,}{z}\right)\times\bv{B}(\omega,z)&=\frac{1}{c^2}\left(\sigma(\omega,z)+i\omega\right)\bv{E}(\omega,z)\\
	\left(i\bvq+\hat{\bv{z}}\dpar{\,}{z}\right)\times\bv{E}(\omega,z)&=-i\omega\bv{B}(\omega,z).
\end{flalign*}
The response of the system is defined by the conductivity $\sigma(\omega,z)$ and, as in the transfer matrix method, we solve for the incident and reflected fields by working backwards from the transmitted field. 
For an angle of incidence $\theta$ the incoming wavevector is $\bvq^I=\frac{\omega}{c}\left[\sin(\theta),0,\cos(\theta)\right]$ and the wavevector of the fields transmitted into the silicon substrate is $\bvq^T=\frac{\omega}{c}\left[\sin(\theta),0,n_{Si}\cos(\theta)\right]$.  Then the ``initial'' conditions of the fields transmitted are, for $s$-polarized light 
\begin{flalign*}
	E^T&=\left[0,1,0\right]E^0\\
	B^T&=\frac{\omega}{c^2}\left[q^T_z,0,-q^T_x\right]E^0
\end{flalign*}
and for $p$-polarized light
\begin{flalign*}
	E^T&=\frac{c}{\omega}\left[-q^T_z,0,q^T_x\right]E^0\\
	B^T&=\frac{1}{c}\left[0,1,0\right]E^0,
\end{flalign*}
where $E^0$ is an arbitrary field strength. In these cases Maxwell's equations reduce to 
\begin{flalign*}
	\dpar{E_y}{z}&=i\omega B_x(z)\\
	\dpar{B_x}{z}&=\left[\frac{1}{c^2}\left(i\omega+\sigma_{yy}(z)\right)-\frac{iq_x^2}{\omega} \right]E_y(z)\\
\end{flalign*}
and
\begin{flalign*}
	\dpar{E_x}{z}&=-\left[i\omega+\frac{c^2q_x^2}{i\omega+\sigma_{zz}(z)}\right]B_y(z)\\
	\dpar{B_y}{z}&=-\frac{1}{c^2}\left(i\omega+\sigma_{xx}(z)\right) E_x(z)\\
\end{flalign*}
for  $s$-polarized and  $p$-polarized light, respectively.  Solving these we obtain the total field, which is easily decomposed into the incident and reflected fields, allowing us to compute the reflection coefficients and ellipsometric spectra.

\section{Sample Structure and Composition}
\begin{figure}
	\centering
	\includegraphics[width=.35\columnwidth]{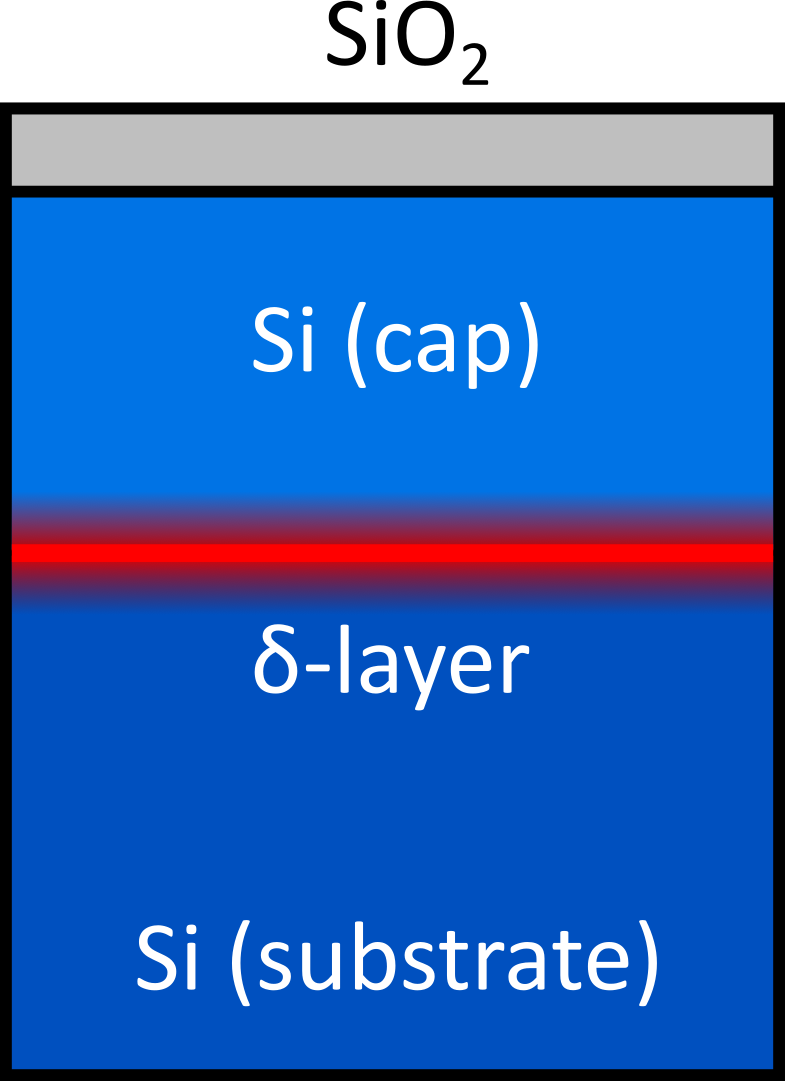}
	\caption{The composition of samples under investigation.  The \dl s (red) are deposited on a silicon substrate (dark blue), upon which an additional layer of silicon (light blue) is deposited.  This ``cap'' frequently contains significant impurity/defect density and will be modeled as such. On top of the stack a thin layer ($\sim 2$nm) of SiO$_2$ (grey) forms, the thickness of which varies slightly from sample to sample.   \label{fig:stack}}
\end{figure}
The samples under consideration have the structure and composition shown schematically in Fig. ~\ref{fig:stack} and were prepared following previously published methods \cite{Ward2017,Katzenmeyer2020,Halsey2022}. A phosphorus \dl~ is deposited by exposing a silicon (100) surface to phosphine in a vacuum chamber and annealing to incorporate the phosphorus. This may be followed by deposition of a locking layer\cite{Keizer2015,Halsey2022}, though the samples in this paper were produced without one.  After this a “capping” layer of epitaxial silicon is deposited on top to activate the dopants, with a thin layer of SiO$_2$ subsequently formed upon exposure to air. Variations in the substrate temperature during silicon epitaxy\cite{Goh_2004} can be used to tune the sharpness of the phosphorus concentration profile.   The epitaxial cap is known to be imperfect, containing disorder and contaminants, but is compatible with device and characterization sample creation. The base is modeled as undoped silicon, while the capping layer is treated as silicon plus a uniform, low density Drude metal. The SiO$_2$ is modeled with a static dielectric constant and we do not attempt to fit the known features originating from phonons.

The overall conductivity function of the stack is then
\begin{flalign*}
	\sigma(\omega,z)&=\sigma_{\rm Si}(\omega,z)+\sigma_{\rm CAP}(\omega,z)+\sigma_{{\rm SiO}_2}(\omega,z)+\sigma_{\rm DL}(\omega,z)\\
	\sigma_{\rm Si}(\omega,z)&=\sigma_{0,{\rm Si}}\frac{\omega_{\rm Si}^2}{\omega^2+\omega_{\rm Si}^2}\theta\left(z-w_{\rm CAP}-w_{{\rm SiO}_2}\right)\left[\begin{array}{ccc}
		1 & 0 & 0\\
		0 & 1 & 0\\
		0 & 0 & 1\\
	\end{array}\right]\\
	\sigma_{\rm CAP}(\omega,z)&=\frac{e^2n_{\rm CAP}}{m_{\rm CAP}}\frac{1}{\omega+ i\frac{1}{\tau}}\left[\theta\left(z-w_{{\rm SiO}_2}\right)-\theta\left(z-w_{\rm CAP}-w_{{\rm SiO}_2}\right)\right]\left[\begin{array}{ccc}
		1 & 0 & 0\\
		0 & 1 & 0\\
		0 & 0 & 1\\
	\end{array}\right]\\
	\sigma_{{\rm SiO}_2}(\omega,z)&=\sigma_{0,{\rm SiO}_2}\left[\theta\left(z\right)-\theta\left(z-w_{{\rm SiO}_2}\right)\right]\left[\begin{array}{ccc}
		1 & 0 & 0\\
		0 & 1 & 0\\
		0 & 0 & 1\\
	\end{array}\right]\\
\end{flalign*}
where $\sigma_{{\rm Si}}$ and $\sigma_{{\rm SiO}_2}$ are the low-frequency conductivities of Si (fit using \dl-free samples due to slight process variation) and SiO$_2$ (based on a assumed relative permittivity of 3.6), respectively, $\omega_{\rm Si}$ is a frequency associated with interband transitions in silicon (also fit using \dl-free samples), $n_{\rm CAP}$ is the carrier density in the capping layer, $m_{\rm CAP}$ is the overall conductivity effective mass of such carriers (we use 0.28 for $n$-type and 0.36 for $p$-type), $w_{\rm CAP}$ and $w_{{\rm SiO}_2}$ are the thicknesses of the two layers, $\tau$ is the relaxation time in the capping layer which we will take to be the same as $\tau$ for \dl s, and $\sigma_{\rm DL}$ refers to a \dl, if present, described by Model I or II. 
\begin{figure}
	\centering
	\subfigure[]{\includegraphics[width=.49\columnwidth]{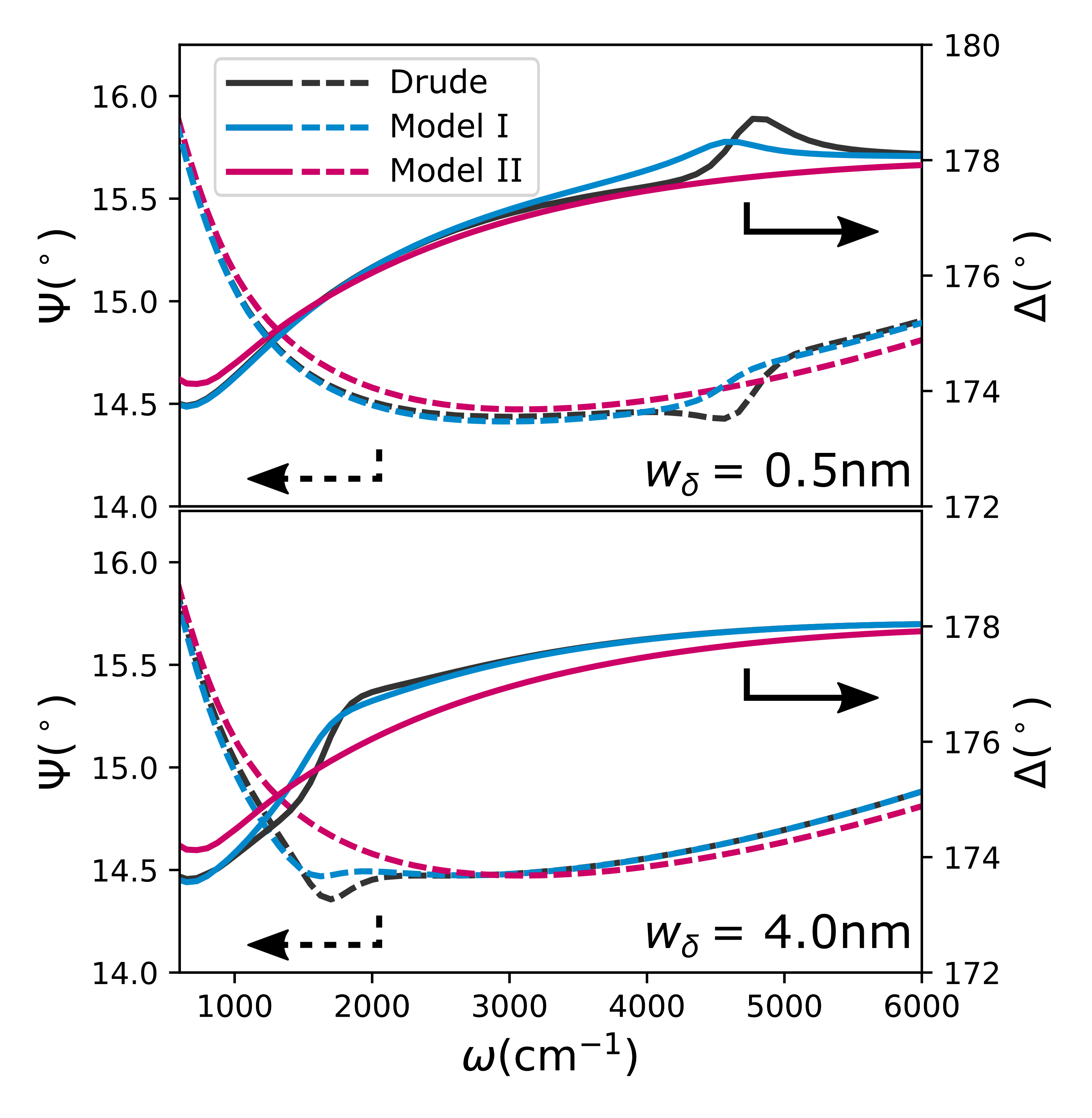}\label{fig:modelcomp}}
	\subfigure[]{\includegraphics[width=.49\columnwidth]{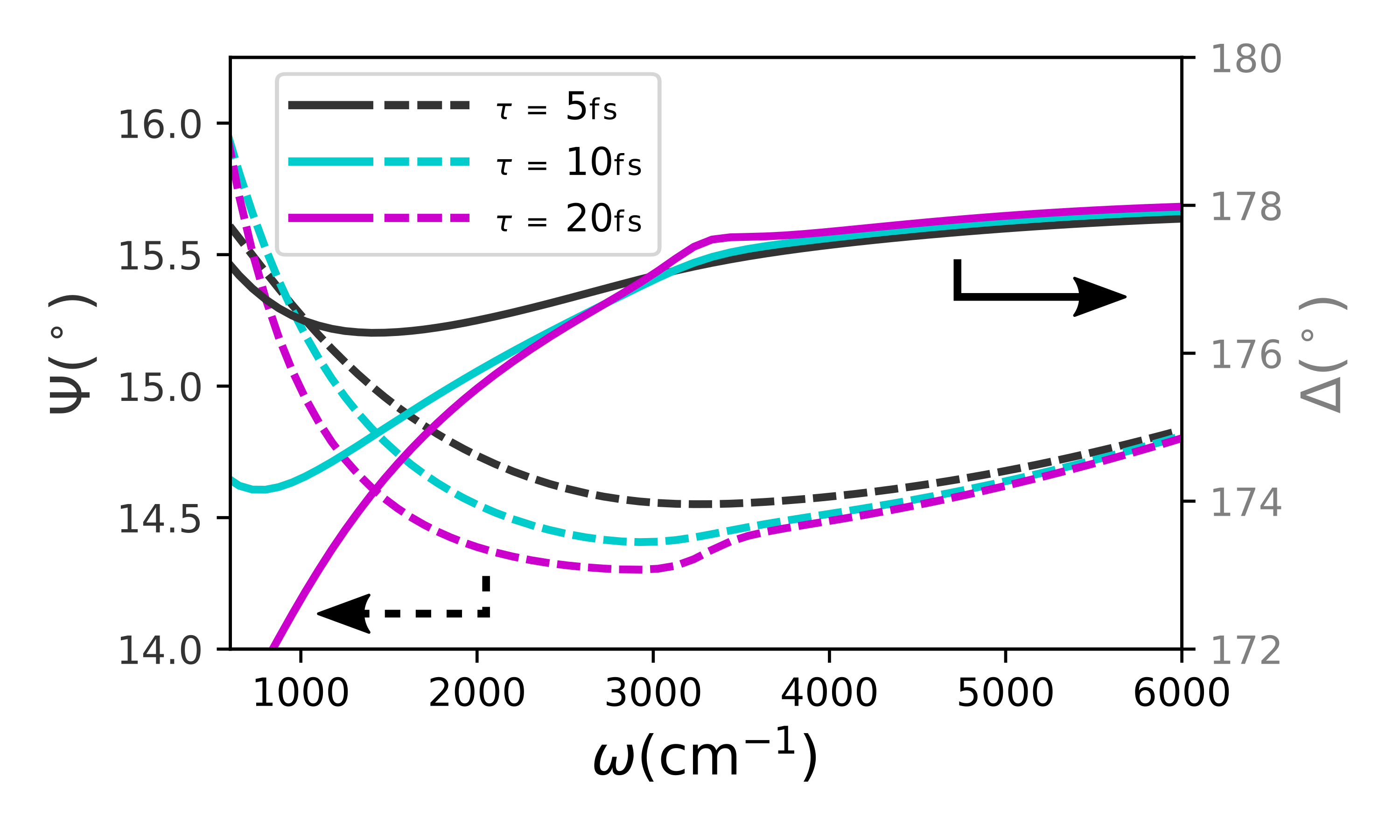}\label{fig:type2-tau3}}
	\caption{ Comparisons of simulations of ellipsometric spectra of hypothetical samples of the kind in Fig.~\ref{fig:stack} containing clean caps and high-density ($10^{14}/$cm$^2$) \dl s. \subref{fig:modelcomp} A classical, isotropic Drude model with a boxcar profile (Drude), Model I, and Model II, for characteristic electronic \dl~widths $w_\delta$ of 0.5nm (top) and 4.0nm (bottom).  \subref{fig:type2-tau3} Model II for varying relaxation times $\tau$ and $w_\delta=0.5$nm.
		\label{fig:tests}}
\end{figure}

In order to understand the impact of the differences in the models, we will first compare the spectra of hypothetical stacks.  We will include for reference a conventional 3D Drude model description of the \dl~(\ie a version of Model I with a boxcar profile for $f(z)$).  We assume in all cases carrier sheet density $n_{2D}=10^{14}/$cm$^{-2}$, $d_{\rm CAP}=50$nm, $d_{{\rm SiO}_2}=2$nm, and $\omega_{\rm Si}=3.5$eV. In Fig.~\ref{fig:modelcomp}, these three models are shown for $\tau=10$fs and varying electronic \dl~thicknesses.

All three models share similar overall spectral features: as the frequency drops below about 2000\cmi, $\Psi$ and $\Delta$ begin to sharply increase and decrease, respectively.  However, the results from these models are distinguishable. In the conventional Drude model, the plasma resonance is clearly visible even for the short relaxation time used, whereas under Model I, due to the smooth envelope function, the plasma resonance is substantially muted, though still distinct. The envelope function modulates the carrier density over the \dl, producing a range of plasma frequencies associated with the \dl. We note that as the envelope becomes wider, the changing density distribution causes the plasma resonance shift to lower frequencies; rather than a peak, the plasma resonance creates an elbow-like feature in both spectra.  In samples well-described by Model I, this will allow determination of electronic \dl~thickness unless $\tau$ is particularly short ($\sim 1$fs). Model II shows a more pronounced effect with a missing plasma resonance feature. In this case the much larger longitudinal effective mass for the out-of-plane conductivity of the $\Gamma$ valley significantly diminishes its contribution, meaning the reflectance of both the $s$- and $p$-polarized light is primarily dependent on the in-plane conductivity.  The ratio $r_p/r_s$ is therefore much less impacted by the plasma resonance, despite producing a clear Drude tail. While the $\Delta$ valley carriers have smaller effective mass in the out-of-plane direction (due to the transverse component), here the much shorter $\tau$ suppresses the plasma resonance feature.  As a result of the lack of plasma resonance, the spectra for Model II are largely insensitive to the thickness (and band energy differences $\omega_\Gamma$) at the given $\tau$; in the following we provide the model fit for this parameter with the understanding that it should not be taken as being particularly precise in this regime.  However, we emphasize that this insensitivity of Model II to thickness is contingent upon the applicability of Model II, and therefore still serves to indicate that the layer is very thin and well-confined.

The plasma resonance feature is not absent in general.  In Fig.~\ref{fig:type2-tau3}  we show Model II for different values of $\tau$ but the same thickness and $\omega_\Gamma$.  We see that longer relaxation times allow a plasma resonance feature to become visible, in addition to sharpening the divergences at low frequencies. Importantly, the combination of depth and shape of the $\Delta$ spectrum strongly constrain $n_{2D}$ and $\tau$. While increases in $n_{2D}$ and $\tau$ can have similar impacts on the $\Delta$ spectrum, they  have opposite impacts on $\Psi$.   This allows us to determine these two parameters quite accurately when the rest of the sample is clean. Thus the low frequency behavior is critical to accurate fitting, as together $\Psi$ and $\Delta$ here strongly constrain the carrier density and relaxation time describing the \dl. Importantly, while longer relaxation times allow for the return of a plasma resonance, it is easy to distinguish from plasma resonance features in the case of Model I. The comparatively shorter relaxation times at which a similar feature appears under Model I causes the divergence in $\Delta$ to be smoothly cut off, while in the case of Model II the sharp decrease in $\Delta$ continues to much lower angles and frequencies.  Thus, the relationship between the low and high frequency behavior is an indicator of the Model that most accurately describes the material.  

This analysis suggests that the suppression of the plasma resonance in ellipsometry is a consequence of the quantum effects of confinement and signals the breakdown of the classical limit of the Drude model, even accounting for a more realistic density profile.  In this sense the suppression of the plasma resonance is a signature of a true \dl, distinguishing it from poorly confined \dl s, which ought to show a plasma resonance feature, even at small $\tau$.

\begin{figure}
	\centering		
	\includegraphics[width=.49\columnwidth]{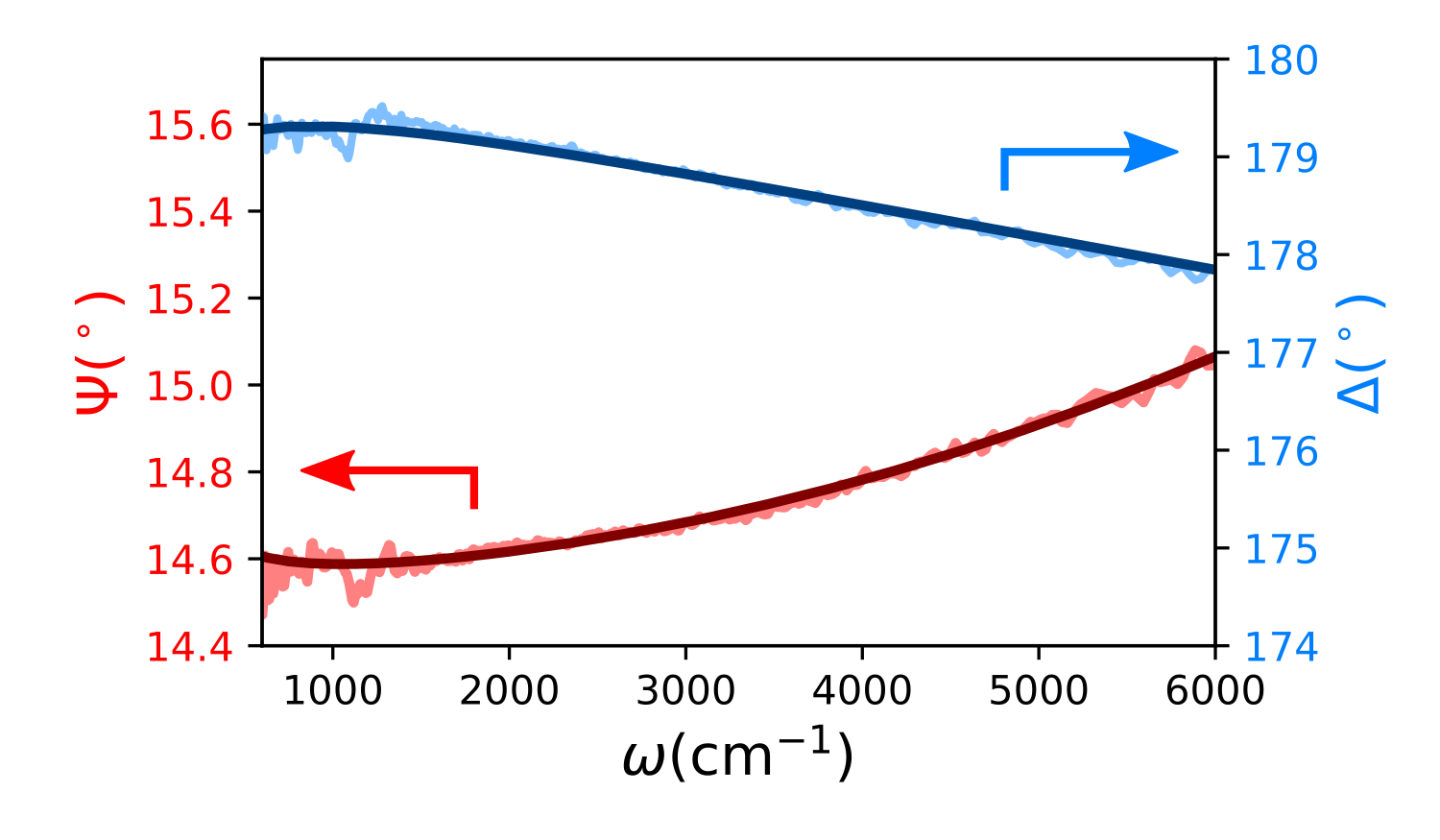}
	\caption{$\Delta$ (blue) and $\Psi$ (red) ellipsometric spectra for a sample with no deposition of a \dl~taken at a 65$^\circ$ angle of incidence. Dark lines correspond to the computed spectra, while noisy, lighter lines are the experimental spectra.  The carrier density in the cap, which is 20nm thick, is found to be $2.5-3.2\times 10^{18}$/cm$^3$ with $\tau=16$fs, and the SiO$_2$ layer is 2.7nm thick. Features around 1000 \cmi are due to vibrational resonances in SiO$_2$}\label{fig:nodelta}
\end{figure}
We now evaluate the success of the models in fitting the elliposometry results for several samples with 1/4ML deposited phosphorus, but different growth conditions for the capping silicon, some of which maintain the sharp doping profile, and others of which promote diffusion.	We first consider a sample with no \dl~to establish a baseline.  No phosphorus was incorporated and a 20 nm cap was deposited 300$^\circ$C at a rate of 0.5nm/min. In Fig.~\ref{fig:nodelta} the experimental ellipsometry results are plotted with the best fit (since there is no \dl, the models produce the same result). $\Psi$ is well-described by bulk silicon with a modest carrier concentration in the cap, while $\Delta$ features a negative slope which can be attributed to the silicon dioxide layer. Due to the low carrier density, there are no plasma resonance features in the frequency range considered. There are deviations of the experimental data from the model between 800\cmi and 1500\cmi and for $\Psi$ that are consistent across samples, though their magnitude varies.  These are reflected in $\Delta$ as well as slightly shifted frequencies.  These are likely due to the impact of SiO$_2$, as they are consistent with known phonon resonances and their magnitude appears to be correlated to its thickness.  While we will not attempt to fit these features, we will deemphasize the response in these frequency ranges in attempting to fit \dl~data. 

\begin{figure}
	\centering
	
	\subfigure[]{\includegraphics[width=.51\columnwidth]{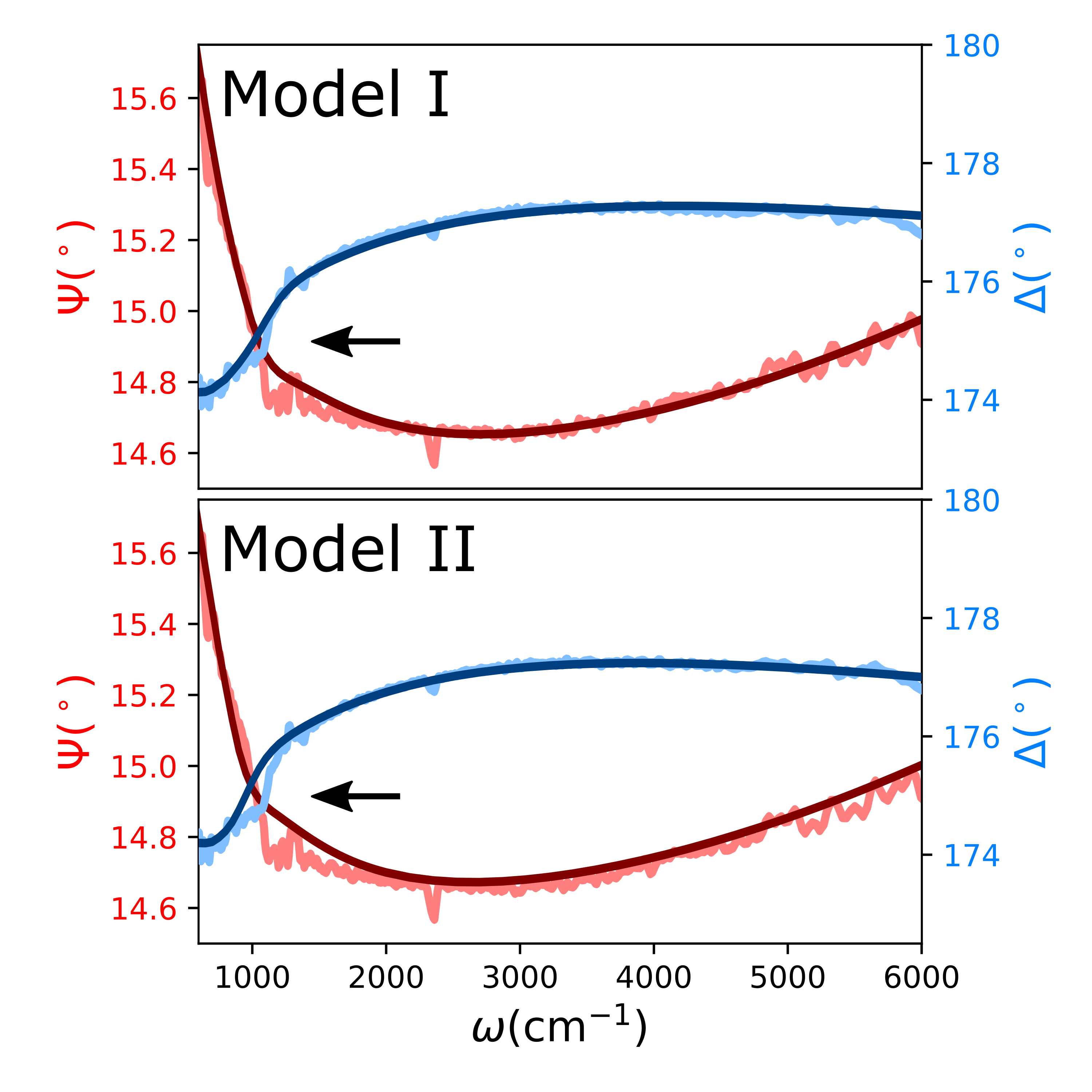}\label{fig:delta1-elps}}
	\subfigure[]{\includegraphics[width=.48\columnwidth]{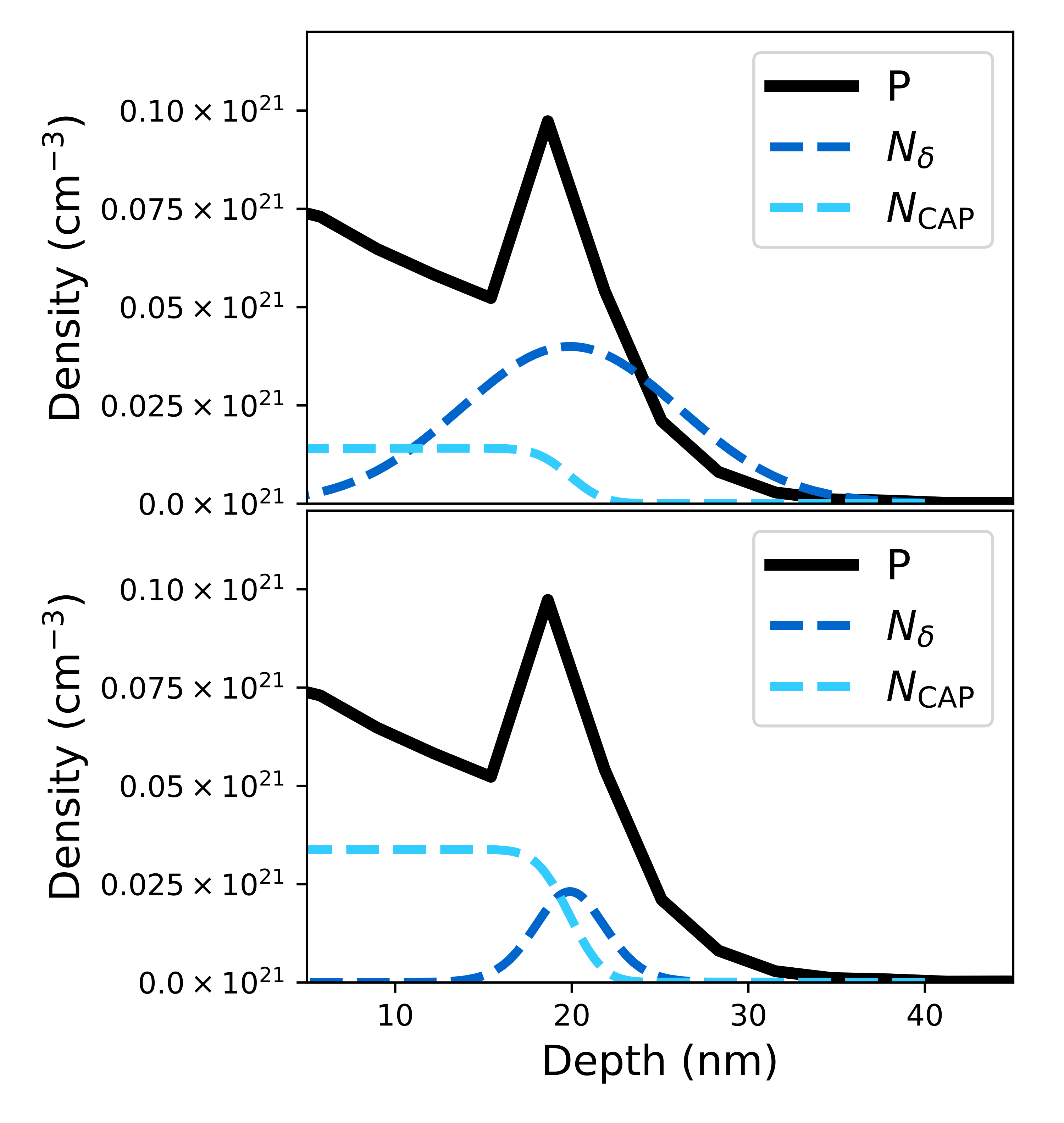}\label{fig:delta1-sims}}
	\caption{\subref{fig:delta1-elps} Elliposometry spectra for a second sample with an intentionally diffused \dl~ -- corresponding to a carrier density of about $1.6\times10^{14}/$cm$^2$) -- taken at a 65$^\circ$ angle of incidence. The fit in the top panel is for Model I and yields  an $n$-type carrier density in a 20nm thick cap of $1.4\times10^{19}/$cm$^3$, a carrier density in a 6nm \dl~of $6.2\times10^{13}/$cm$^2$, $\tau=11.0$fs, and an SiO$_2$ thickness of 2.7nm. In the bottom panel, the fit to Model II yields  a carrier density in the 20nm thick cap of $3.8\times10^{19}/$cm$^3$ (assuming $n$-type carriers), a carrier density in a 0.9nm \dl~($\omega_\Gamma=50$meV) of $1.2\times10^{13}/$cm$^2$, $\tau=10.4$fs, and an SiO$_2$ thickness of 2.8nm.  Arrows indicate the region where the ``elbows'' appear in the spectra. \subref{fig:delta1-sims} The SIMS analysis for sample showing densities of phosphorus along with Model I and Model II fits for the carrier densities in the \dl~and cap.  Due to the resolution limit of the SIMS analysis, the fitted densities have been convolved with a Gaussian (FWHM=3.5nm) to aid comparison.  }\label{fig:delta1}
\end{figure}
In Fig.~\ref{fig:delta1} we show the data for a sample where dopant incorporation and deposition of the cap layer has been done in a way that promotes adatom-mediated diffusion, and spreads the 1/4 ML of phosphorus throughout the cap during growth. In this case phosphorus was incorporated at 300$^\circ$C, which is below the temperature at which phosphorus is optimally incorporated\cite{Wilson_2006}, thus leaving excess phosphorus to segregate during cap growth. A 20 nm cap was deposited 300$^\circ$C at a rate of 0.2nm/min, with this relatively slow growth rate selected to provide time for adatom-mediated diffusion. Here, the 3D doping profile has resulted in the appearance of a distinct elbow-like feature at low frequency in the ellipsometry data (Fig.\ref{fig:delta1-elps}) - the plasma resonance. Model I with a moderately thick \dl~(6nm) and substantial $n_{\rm CAP}$ is able to fit the data quite well, reproducing the elbows of both $\Psi$ and $\Delta$, and indicating significant diffusion of carriers away from the intended \dl.  The notable deviations at low wavenumber can be attributed to the SiO$_2$-originating signal~\cite{Boher2004} seen in the \dl-free sample. The best fit for Model II yields a very sparse \dl; since a \dl~under Model II cannot provide an elbow-like feature, this spectral feature must be generated almost entirely by a uniform carrier density in the cap.  The fit is nonetheless still fairly good, which is not particularly surprising since the two cases are actually quite similar physically: the overall carrier density profile in Model I contains a significant uniform contribution from the cap and the suggested \dl~thickness means that it is diffused over a considerable portion of the cap.  This is borne out by the SIMS analysis (Fig.~\ref{fig:delta1-sims}), which shows significant phosphorus density in the cap in addition to a broad peak at the depth of the intended \dl.  Overall the modeling accurately reflects the samples nature as a poorly confined dopant-layer.  For both models the total sheet density from the cap and \dl~is $\sim0.9\times10^{14}/$cm$^2$, about half of the intended \dl~density. 

\begin{figure}
	\centering
	\subfigure[]{\includegraphics[width=.51\columnwidth]{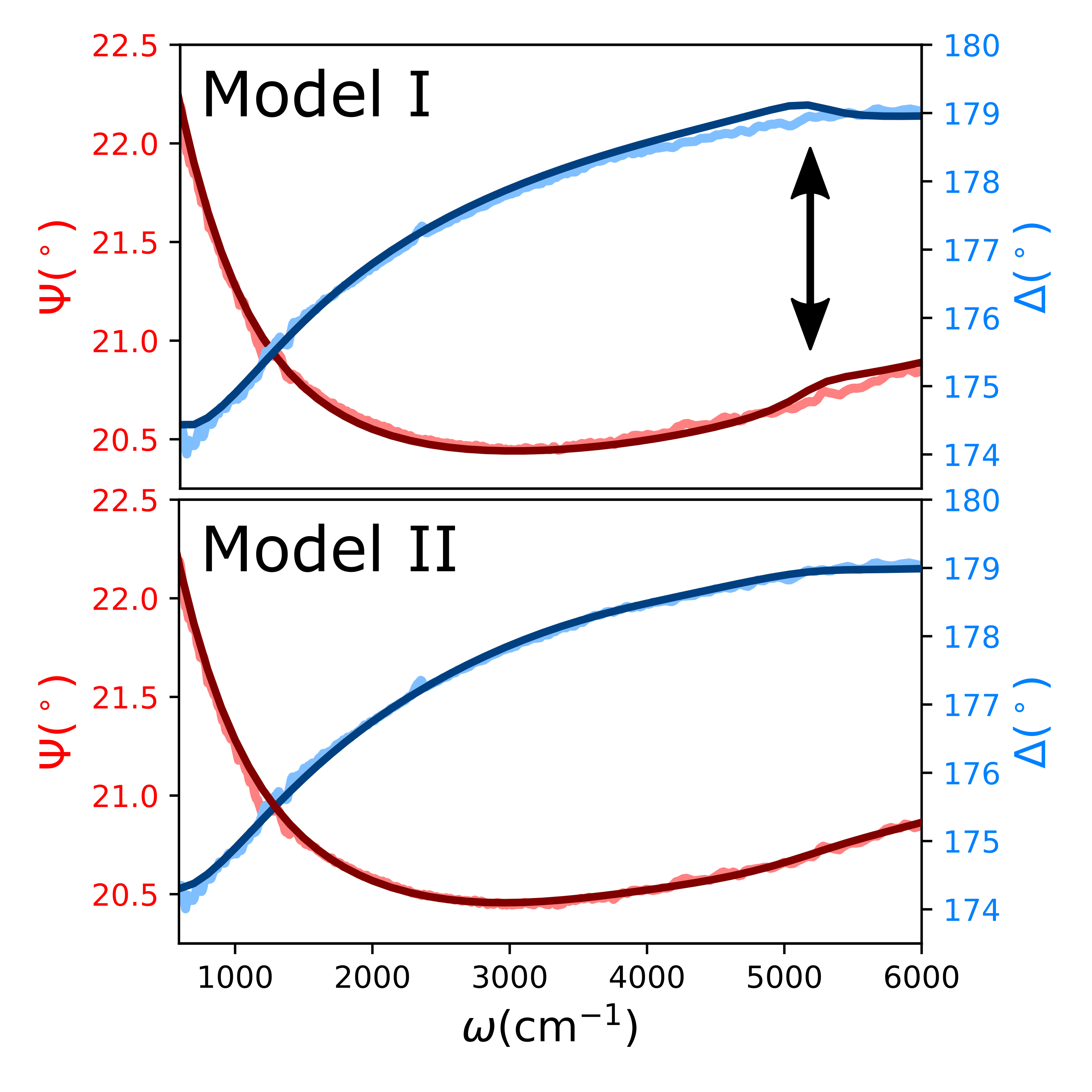}\label{fig:delta3-elps}}
	\subfigure[]{\includegraphics[width=.48\columnwidth]{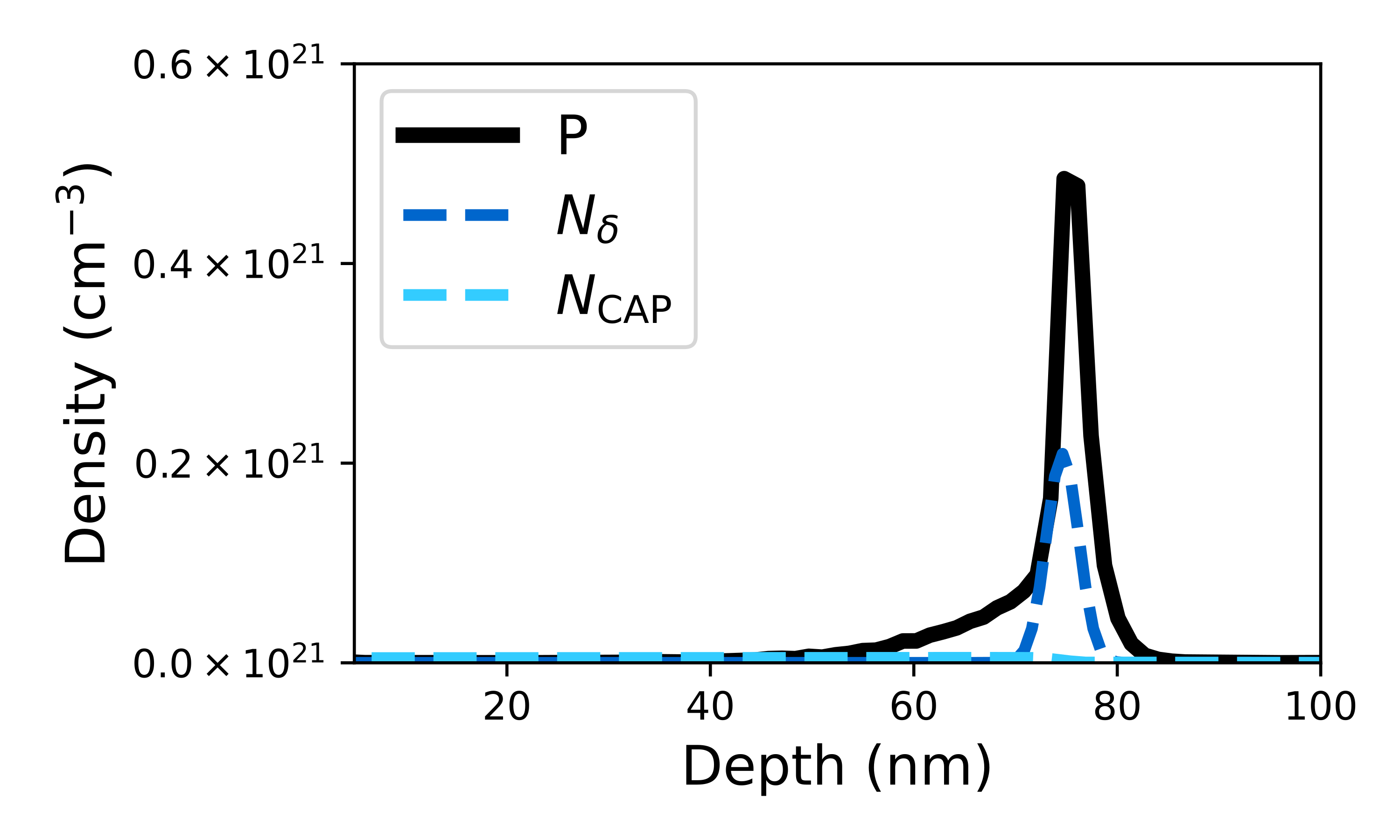}\label{fig:delta3-sims}}
	
	\caption{\subref{fig:delta3-elps} 	Elliposometry spectra for a third sample with an intended $1/4$ML \dl~taken at a 60$^\circ$ angle of incidence. The fit in the top panel is for Model I and carrier density in a 75nm thick cap of $2\times10^{19}/$cm$^3m_{\rm CAP}$, a carrier density in a 0.4nm \dl~of $1.04\times10^{14}/$cm$^2$, $\tau=10.1$fs, and an SiO$_2$ thickness of 2.1nm. A double arrow indicates the expected location of the plasma resonance feature. In the bottom panel, the fit to Model II yields the same carrier density in the 75nm thick cap ($5.6-7.2\times10^{18}/$cm$^3$), a carrier density in a 0.26nm \dl~($\omega_\Gamma=550$meV) of $8.8\times10^{13}/$cm$^2$, $\tau=12.8$fs, and an SiO$_2$ thickness of 2.1nm.
		\subref{fig:delta3-sims} The SIMS analysis for sample showing densities of phosphorus along with fitted densities for Model II of carriers in the \dl~and cap, assuming $p$-type carriers in the latter. }\label{fig:delta3}
\end{figure}

Incorporating dopants and growing the capping silicon under different conditions can produce \dl s with significantly tighter confinement of the dopant atoms. For this sample phosphorus was incorporated at 380$^\circ$C, which is near the temperature at which phosphorus is optimally incorporated \cite{Wilson_2006}, thus minimizing excess phosphorus available for segregation during cap growth.  A 75nm cap was deposited at 300$^\circ$C at a relatively fast rate of 0.75nm/min to suppress adatom-mediated diffusion. In Fig.~\ref{fig:delta3-elps} we see spectra with no obvious elbow at low frequency, suggesting a much narrower \dl~with reduced population of the $\Delta$ valleys. Fits for both models corroborate this, clearly indicating a subnanometer carrier distribution and a well-confined \dl. However, Model I produces a distinct plasma resonance feature in both $\Psi$ and $\Delta$ at $\sim 5000$\cmi.  Model II correctly results in  smooth spectra in this regime, matching the experimental spectra extremely well.  The superior fit of Model II is consistent with the notion that the anisotropy introduced by confinement effects is necessary to describe well-confined \dl s. Both models indicate some carrier density in the cap, consistent with other results using this process.   The SIMS analysis (Fig.~\ref{fig:delta3-sims}) corroborates this, showing a very narrow peak in the phosphourous density that is matched by the profile of the Model II fit. The cap carrier density, which is comparable to the sample with no phosphorus doping in Fig. 4, primarily originates from unintended aluminum dopants~\cite{Anderson_2020}, with some contribution from the phosphorus tail of the \dl.  The carrier density of the \dl~yielded by the model is again about half of the intended density. We note that the density obtained from Hall measurements is around $1.8\times10^{14}/$cm$^2$, consistent with the targeted carrier density and suggesting that the ellipsometry is systematically underestimating the carrier density by a factor of about two.  This can be considered good agreement for ellipsometry, and the systematic nature of the error could allow for accurate estimation of carrier densities from ellipsometry.

\section{Conclusion}
We have derived a physically realistic model for the AC conductivity of P:\dl s, and used it to model the ellipsometry spectra of multiple samples containing  \dl s of varying thickness by taking into consideration the quantum confinement of carriers in the P:\dl.  We obtain very good fits, resolving the previously unexplained lack of plasma resonance features that would be expected from a classical Drude model, which is due to the sharpness of the charge density profile and valley degeneracy producing an effective mass and relaxation time anisotropy.  As such, the suppression of the plasma resonance -- to the point of elimination in the considered samples -- serves as a signature of well-confined \dl s and two-dimensional physics. As a \dl~widens and diffuses out, essentially ceasing to be a \dl, confinement effects weaken and the plasma resonance feature reappears as an elbow-like feature in the spectra, with its position roughly corresponding to the width of the \dl.   Finally, we predict that doubling the relaxation time in well-confined \dl s will cause the plasma resonance to reappear. In combination with a very sharp Drude tail associated with longer relaxation times, this would provide a unique signature of well-confined \dl~that cannot be reproduced by a Drude metal type layer.  The reappearance of this plasma feature would furnish additional information about the \dl's electronic structure, much as it does in much thicker layers having a plasma resonance. Our demonstration that ellipsometry spectra can be effectively modeled and used to non-destructively assess P:\dl~quality and characteristics can be applied as part of a general approach to suppress or enhance plasma resonances in engineered thin films as desired.

\section{Acknowledgments}
We gratefully acknowledge helpful conversations with Andrew Baczewski and Paul Davids. This work was supported by the Laboratory Directed Research and Development Program at Sandia National Laboratories under project 213017 and was performed, in part, at the Center for Integrated Nanotechnologies, a U.S. DOE, Office of Basic Energy Sciences user facility. This article has been authored by an employee of National Technology \& Engineering Solutions of Sandia, LLC under Contract No. DE-NA0003525 with the U.S. Department of Energy (DOE). The employee owns all right, title and interest in and to the article and is solely responsible for its contents. The United States Government retains and the publisher, by accepting the article for publication, acknowledges that the United States Government retains a non-exclusive, paid-up, irrevocable, world-wide license to publish or reproduce the published form of this article or allow others to do so, for United States Government purposes. The DOE will provide public access to these results of federally sponsored research in accordance with the DOE Public Access Plan https://www.energy.gov/downloads/doe-public-access-plan This paper describes objective technical results and analysis. Any subjective views or opinions that might be expressed in the paper do not necessarily represent the views of the U.S. Department of Energy or the United States Government.

\bibliography{ellips}
\end{document}